# Formation of Bright Solitons from Wave Packets with Repulsive Nonlinearity

Zihui Wang,[1] Mikhail Cherkasskii,[2] Boris A. Kalinikos,[2] Lincoln D. Carr,[3] and Mingzhong Wu[1]*

[1]*Department of Physics, Colorado State University, Fort Collins, Colorado 80523, USA*

[2]*St.Petersburg Electrotechnical University, 197376, St.Petersburg, Russia*

[3]*Department of Physics, Colorado School of Mines, Golden, Colorado 80401, USA*

Formation of bright envelope solitons from wave packets with a repulsive nonlinearity was observed for the first time. The experiments used surface spin-wave packets in magnetic yttrium iron garnet (YIG) thin film strips. When the wave packets are narrow and have low power, they undergo self-broadening during the propagation. When the wave packets are relatively wide or their power is relatively high, they can experience self-narrowing or even evolve into bright solitons. The experimental results were reproduced by numerical simulations based on a modified nonlinear Schrödinger equation model.

Solitons are a universal phenomenon in nature, appearing in systems as diverse as water, optical fibers, electromagnetic transmission lines, deoxyribonucleic acid, and ultra-cold quantum gases.[1,2,3,4,5] The formation of solitons from large-amplitude waves can be described by paradigmatic nonlinear equations, one of which is the nonlinear Schrödinger equation (NLSE). In the terms of the NLSE model, two classes of envelope solitons, bright and dark, can be excited in nonlinear media. A bright envelope soliton is a localized excitation on the envelope of a large-amplitude carrier wave. It typically takes a hyperbolic secant shape and has a constant phase across its width.[6] A dark envelope soliton is a dip or null in a large-amplitude wave background. When the dip goes to zero, one has a black soliton. When the amplitude at the dip is nonzero, one has a gray soliton. A dark soliton has a jump in phase at its center. For a black soliton, such a phase jump equals to ´. For a gray soliton, the phase jump is between 0 and ´. The envelope of a dark soliton can be described by a unique function.[3] For a black soliton, this function is typically a hyperbolic tangent function.

According to the NLSE model, the formation of a bright soliton from a large-amplitude wave packet is possible in systems with an attractive (or self-focusing) nonlinearity and is prohibited in systems with a repulsive (or defocusing) nonlinearity. The underlying physics is as follows. The attractive nonlinearity produces a pulse self-narrowing effect; at a certain power level the self-narrowing can balance the dispersion-induced pulse self-broadening and give rise to the formation of a bright envelope soliton. In contrast, in systems with a repulsive nonlinearity the nonlinearity induces self-broadening of the wave packet, just as the dispersion does, and thereby disables the formation of a bright soliton. Previous experiments show good agreements with these theoretical predictions: the formation of bright solitons from wave packets has been demonstrated in different systems with an attractive nonlinearity,[3,7] while the self-broadening has been observed for wave packets in systems with a repulsive nonlinearity.[8]

This letter reports on the first observation of the formation of bright solitons from wave packets with a repulsive nonlinearity. The experiments made use of spin waves traveling along long and narrow magnetic yttrium iron garnet ($Y_3Fe_5O_{12}$, YIG)[9] thin film strips. The YIG strips were magnetized by static magnetic fields applied in their planes and perpendicular to their length directions. This film/field configuration supports the propagation of surface spin waves with a repulsive nonlinearity.[10,11] To excite a spin wave packet in the YIG strip, a microstrip line transducer was placed on one end of the YIG strip and was fed with a microwave pulse. As the spin wave packet propagates along the YIG strip, it was measured by either a secondary microstrip line or a magneto-dynamic inductive probe located above the YIG strip. When the input microwave pulse is relatively narrow and has relatively low power, one observes the broadening of the spin wave packet during its propagation. At certain large input pulse widths and high power levels, however, the spin wave packet undergoes self-narrowing and evolves into a bright envelope soliton. The formation of this soliton is contradictory to the prediction of the standard NLSE model, but was reproduced by numerical simulations with a modified NLSE model that took into account damping and saturable nonlinearity.

Figure 1 shows representative data on the formation of bright solitons from surface spin-wave packets. Graph (a) shows the experimental configuration. The YIG film strip was cut from a 5.6-μm-thick (111) YIG wafer grown on a gadolinium gallium garnet substrate. The strip was 30 mm long and 2 mm wide. The magnetic field was set to 910 Oe. The input and output transducers were 50-μm-wide striplines and were 6.3 mm apart. The input microwave pulses had a carrier frequency of 4.51 GHz. Note that, in Fig. 1 and other figures as well as the discussions below, $P_{in}$ denotes the nominal microwave pulse power applied to the input transducer, $\tau_{in}$ denotes the half-power width of the input microwave pulse, $P_{out}$ is the power of the output signal, and $\tau_{out}$ represents the half-power width of the output pulse. In Fig. 1, graphs (b), (c), (d), (f), and (g) give the power profiles of the output signals measured with different $P_{in}$ and $\tau_{in}$ values, as indicated. The circles in (d) shows a fit to the hyperbolic secant squared function.[1,3] Graph (e) shows the corresponding phase ($\theta$) profile of the signal shown in (d). Here, the profile shows the phase relative to a reference continuous wave whose frequency equals to the carrier frequency of the input microwave.[6] Graph (h) shows the change of $\tau_{out}$ with $P_{in}$ for a fixed $\tau_{in}$, as indicated, while graph (i) shows the change of $\tau_{out}$ with $\tau_{in}$ for a fixed $P_{in}$, as indicated.

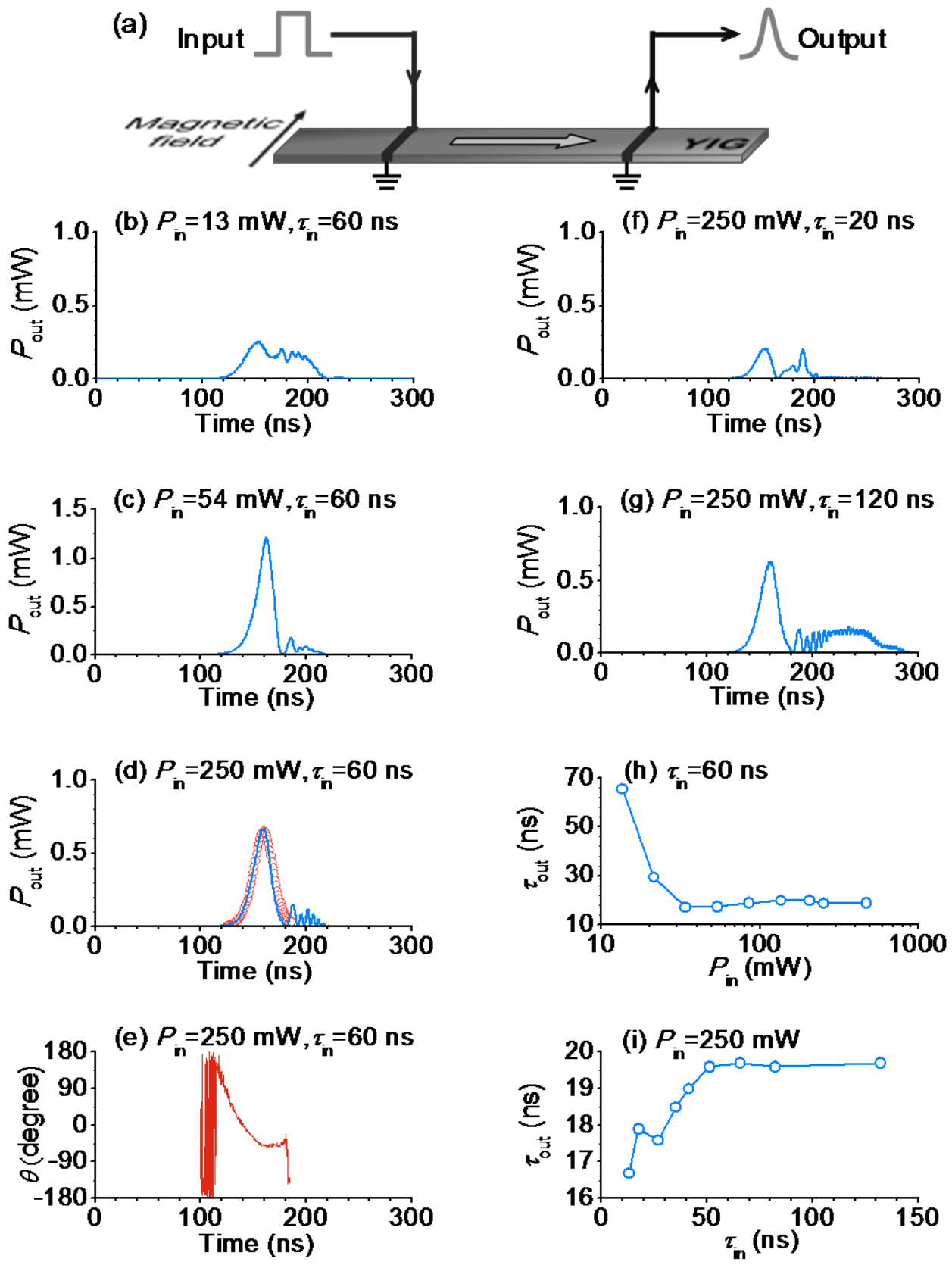

FIG. 1. Propagation of spin-wave packets in a 2.0-mm-wide YIG strip. (a) Experimental setup. (b), (c), (d), (f), and (g) Envelopes of output signals obtained at different input pulse power levels ($P_{in}$) and widths ($\tau_{in}$). (e) Phase ($\theta$) profile for the signal shown in (d). (h) Width of output pulse ($\tau_{out}$) as a function of $P_{in}$. (i) Width of output pulse as a function of $\tau_{in}$.

The data in Fig. 1 show three important results. (1) The data in Figs. 1 (b)-(e) and (h) show the change of the output signal with the input power $P_{in}$. One can see that the output pulse is broader than the input pulse when $P_{in}$=13 mW, as shown in (b), and is significantly narrower when $P_{in}$>30 mW, as shown in (c), (d), and (h). This indicates that the spin-wave packet undergoes self-broadening at low power and self-narrowing at relatively high power. (2) The data in Figs. 1 (d), (f), (g), and (i) show the change of the output signal with the input pulse width $\tau_{in}$. It is evident that the width of the output pulse increases with $\tau_{in}$ when $\tau_{in}$<50 ns and then saturates to about 19.5 ns when $\tau_{in}$>50 ns. These results indicate that the spin-wave packet experiences strong self-narrowing when it is relatively broad. (3) The pulses shown in (d) and (g) are indeed bright solitons. As shown representatively in (d) and (e), they have a hyperbolic secant shape and a constant phase profile at their centers, which are the two key signatures of bright solitons.[1,6]

The data from Fig. 1 clearly demonstrate the formation of bright solitons from surface spin-wave packets when the energy of the initial signals (the product of $P_{in}$ and $\tau_{in}$) is beyond a certain level. This result is contradictory to the predictions of the NLSE model. One possible argument is that the width of the YIG strip might play a role in the observed formation of bright solitons. To rule out this possibility, similar measurements were carried out with an YIG strip that is an order of magnitude narrower. The main data are as follows.

Figure 2 gives the data measured with a 0.2-mm-wide YIG strip. This figure is shown in the same format as in Fig. 1. In contrast to the data in Fig. 1, the data here were measured by a 50-Ω inductive probe,[12] rather than a secondary microstrip transducer. The distance between the input transducer and the inductive probe was about 2.6 mm. The magnetic field was set to 1120 Oe. The input microwave pulse had a carrier frequency of 5.07 GHz.

The data in Fig. 2 show results very similar to those shown in Fig. 1. Specifically, the low-power, narrow spin-wave packets undergo self-broadening, as shown in (b), (c), (f), and (h); as the power and width are increased to certain levels, the spin-wave packets experience self-narrowing, as shown in (h) and (i), and can also evolve into solitons, as shown in (d), (e), and (g). Therefore, the data in Fig. 2 clearly confirm the results from Fig. 1. This indicates that the formation of solitons reported here is not due to any effects associated with the YIG strip width. Note that the solitons shown in Fig. 2 are narrower than those shown in Fig. 1. This difference results mainly from the fact that the spin-wave amplitudes and dispersion properties were different in the two experiments. The spin-wave dispersion differed in the two experiments because the magnetic fields were different and the wave numbers of the excited spin-wave modes were also not the same.

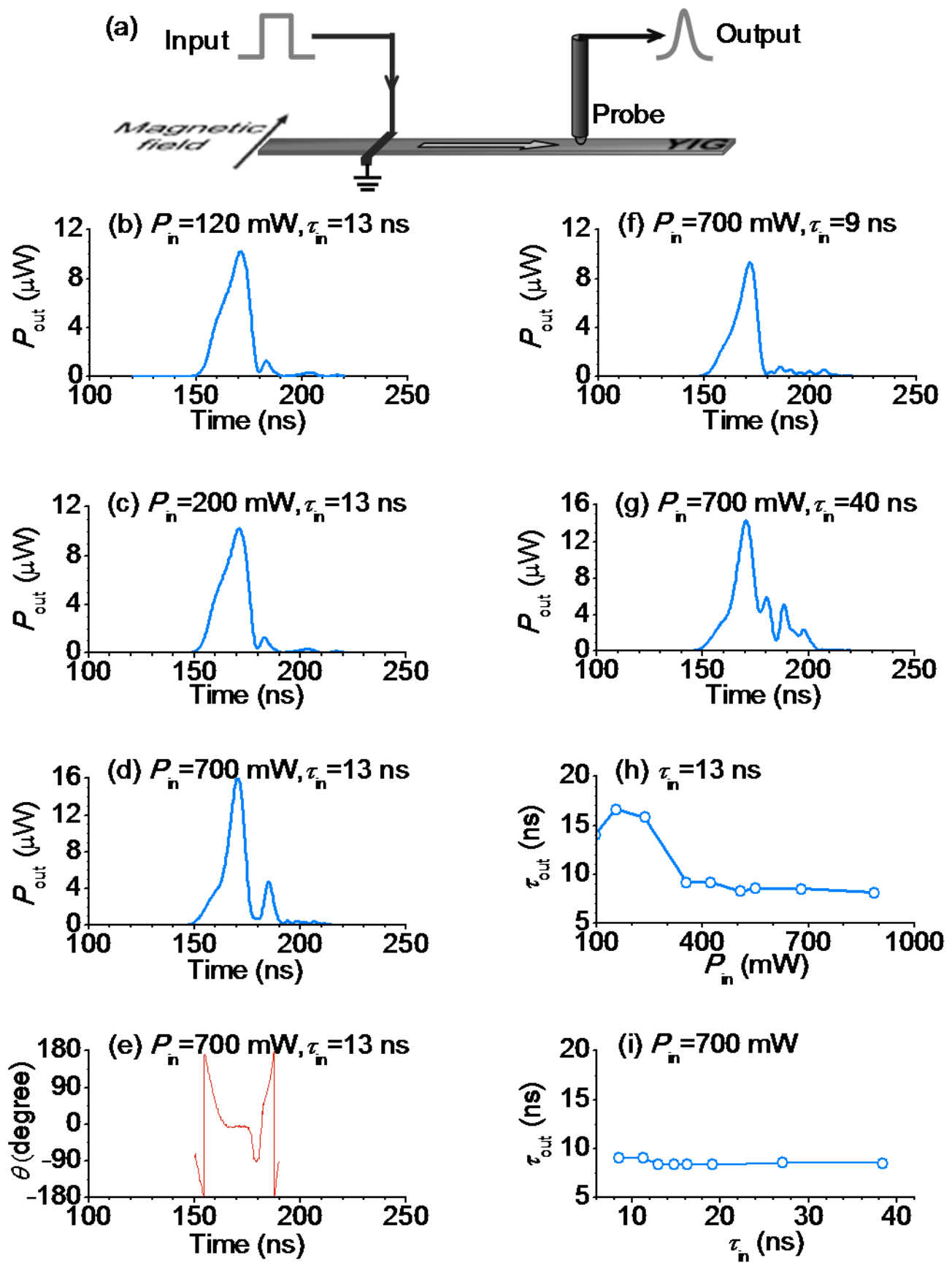

FIG. 2. Propagation of spin-wave packets in a 0.2-mm-wide YIG strip. (a) Experimental setup. (b), (c), (d), (f), and (g) Envelopes of output signals obtained at different input pulse power levels ($P_{in}$) and widths ($\tau_{in}$). (e) Phase ($\theta$) profile for the signal shown in (d). (h) Width of output pulse ($\tau_{out}$) as a function of $P_{in}$. (i) Width of output pulse as a function of $\tau_{in}$.

Turn now to the spatial formation of solitons from surface spin-wave packets. Figure 3 shows representative data. Graph (a) gives the profile of an input signal. The power and carrier frequency of the input signal were 700 mW and 5.07 GHz, respectively. Graphs (b)-(f) give the corresponding output signals measured with the same experimental configuration as depicted in Fig. 2(a). The signals were measured by placing the inductive probe at different distances ($x$) from the input transducer, as indicated. The red curves in (b)-(f) are the corresponding phase profiles.

The data in Fig. 3 show the spatial evolution of a spin-wave packet. At $x$=1.1 mm, the packet has a width similar to that of the input pulse. As the packet propagates to $x$=2.1 mm, it develops into a soliton, which is not only much narrower than both the initial pulse and the packet at $x$=1.1 mm but also has a constant phase at its center portion, as shown in (c). At $x$=2.6 mm, the packet has a lower amplitude due to the magnetic damping but still maintains its solitonic nature, as shown in (d). As the packet continues to propagate further, it loses its solitonic properties and undergoes self-broadening, as shown in (e) and (f), due to significant reduction in amplitude. Note that the phase profiles for all the signals in (b), (e), and (f) are not constant. These results support the above-drawn conclusion, namely, that it is possible to produce a bright soliton from a surface spin-wave packet.

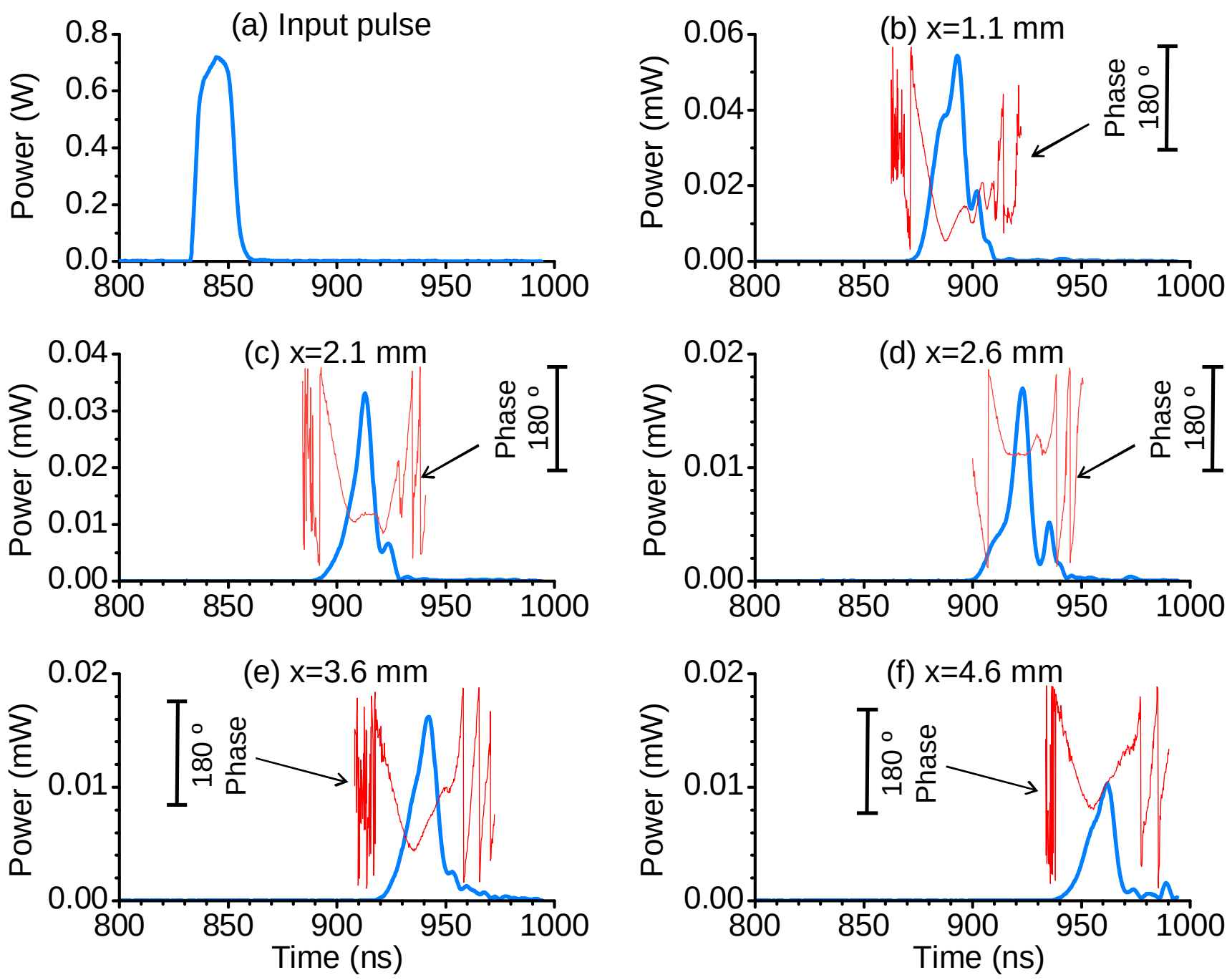

FIG. 3. Spatial formation of a spin-wave soliton in a 0.2-mm-wide YIG strip. (a) Profile of an input signal. (b)-(f) Profiles of output signals measured by an inductive probe placed at different distances ($x$) from the input transducer. The red curves in (b) and (f) are the corresponding phase profiles.

The data in Fig. 3 also indicate the other two important results. (1) The development of a soliton takes a certain distance, about 2 mm for the above-cited conditions, due to the fact that the nonlinearity effect needs a certain propagation distance to develop. (2) The soliton exists only in a relatively short range, about 1-2 mm for the above-cited conditions, due to the damping of carrier spin waves. To increase the "life" distance or lifetime of a spin-wave soliton, one can take advantage of parametric pumping[13] or active feedback[9] techniques.

As mentioned above, the soliton formation presented here is contradictory to the standard NLSE model. However, it can be reproduced by numerical simulations based on the equation

$$i\left[\frac{\partial u}{\partial t}+v_g\frac{\partial u}{\partial x}+\eta u\right]-\frac{1}{2}D\frac{\partial^2 u}{\partial x^2}+\left(N|u|^2+S|u|^4\right)u=0 \tag{1}$$

where $u$ is the amplitude of a spin-wave packet, $x$ and $t$ are spatial and temporal coordinates, respectively, $v_g$ is the group velocity, $\eta$ is the damping coefficient, $D$ is the dispersion coefficient, and $N$ and $S$ are the cubic and quintic nonlinearity coefficients, respectively. The quantic nonlinearity term is included because the cubic nonlinearity is insufficient to capture the experimental observations presented above. This additional term is an expansion to the lowest order of saturable nonlinearity. The simulations used the split-step method to solve the derivative terms with respect to $x$ and used the Runge-Kutta method to solve the equation with the rest of the terms.[14,15] A high-order Gaussian profile was taken in simulations for the input pulse because it is much closer to the experimental situation than a squared pulse. The use of a square pulse as in the input pulse gave rise to numerical noise due to the discontinuity at the pulse's edges. The use of a fundamental Gaussian function did not onsiderably change the simulation results. It should be noted that both the standard and modified NLSE models are for nonlinear waves in one-dimensional (1D) systems, and previous work had demonstrated the feasibility of using the 1D NLSE models to describe nonlinear spin waves in quasi-1D YIG film strips.[16,17]

Figure 4 shows representative results obtained for different initial pulse amplitudes ($u_0$), as indicated. In each panel, the left and right diagrams show the power and phase profiles, respectively. The simulations were carried out for a 20-mm-long 1D film strip and a total propagation time of 250 ns. The film strip was split into 9182 steps, and the temporal evolution step was set to 0.05 ns. The input pulse was a high-order Gaussian profile with an order number of 20 and a half-power width of 15 ns. The other parameters used are as follows: $v_g$=3.8×10$^6$ cm/s, $\eta$=3.1×10$^6$ rad/s, $D$=-4.7×10$^3$ rad·cm$^2$/s, $N$=-10.1×10$^9$ rad/s, and $S$=1.8×10$^{12}$ rad/s. Among these parameters, $v_g$, $D$, $\eta$, and $N$ were calculated according to the properties of the YIG film,9 and the $S$ was optimized for the reproduction of the experimental responses.

The profiles in Fig. 4 indicate that, at low initial power, the pulse is broader than the initial pulse and has a phase profile which is not constant at the pulse center, as shown in (a) and (b); at relatively high power, however, the pulse is not only significantly narrower than the initial pulse but also has a constant phase across its center portion, as shown in (c). These results agree with the experimental results presented above.

The reproduction of the experimental responses with the modified NLSE model indicates the underlying physical processes for the formation of bright solitons from surface spin-wave packets. In comparison with the standard NLSE, the additional terms in the modified equation are $\eta u$ and $S|u|^4 u$. The term $\eta u$ accounts for the damping of spin waves in YIG films, while the term $S|u|^4 u$ is needed for the reproduction of the experimental responses. Since the sign of $S$ was opposite to that of $N$, the term $S|u|^4 u$ played a role opposite to $N|u|^2 u$ and caused nonlinearity saturation. In particular, for the configuration cited for Fig. 4(c) the term $S|u|^4 u$ overwhelmed the term $N|u|^2 u$, resulting in a repulsive-to-attractive nonlinearity transition and the formation of a bright soliton. Thus, one can see that the saturable nonlinearity played a critical role in the formation of the bright solitons from surface spin-wave packets. It should be noted that the saturable nonlinearity has been known as a critical factor for the formation of solitons in optical fibers.[18]

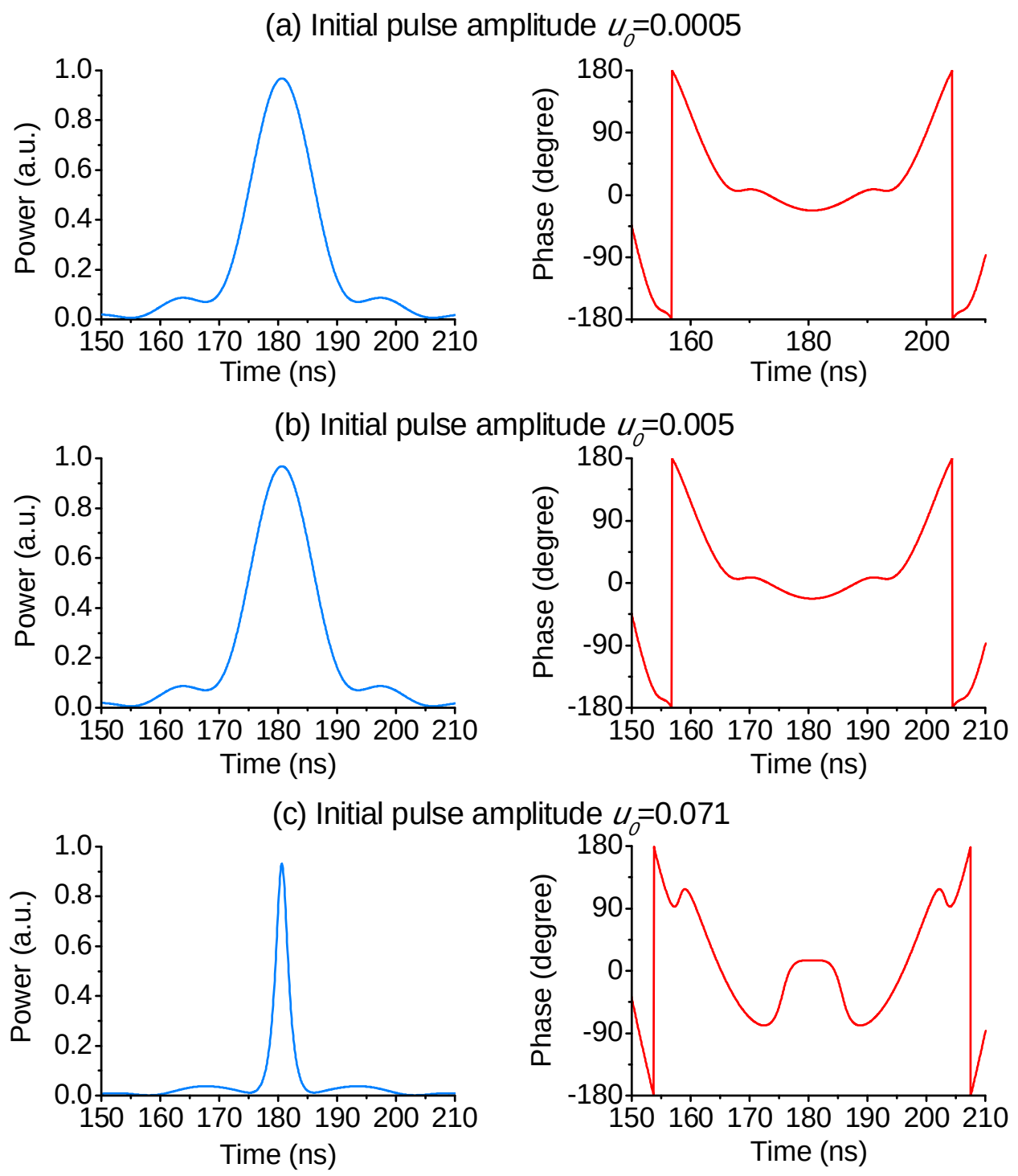

FIG. 4. Power (left) and phase (right) profiles of spin-wave packets propagating in a YIG strip. The profiles were obtained from simulations with different initial pulse amplitudes, as indicated, for a propagation distance of 4.9 mm.

In summary, this letter reports the first observation of the formation of bright solitons from surface spin-wave packets propagating in YIG thin films. The formation of such solitons was observed in YIG film strips with significantly different widths. The spatial evolution of the solitons was measured by placing an inductive probe at different positions along the YIG strip. The experimental observation was reproduced by numerical simulations based on a modified NLSE model. The agreement between the experimental and numerical results indicates that the saturable nonlinearity played important roles in the soliton formation.

This work was supported in part by U. S. National Science Foundation (DMR-0906489 and ECCS-1231598) and the Russian Foundation for Basic Research.

*Corresponding author.

E-mail: mwu@lamar.colostate.edu